\documentclass[aps,prd,longbibliography,superscriptaddress,twocolumn,nofootinbib,10pt]{revtex4-2}
\usepackage{graphicx, comment}
\usepackage{amsmath}
\usepackage{amsfonts}
\usepackage{dsfont}
\usepackage{amssymb}
\usepackage{ifthen}
\usepackage{ulem}
\usepackage{color}
\usepackage{float}
\usepackage{physics}
\usepackage{hyperref}
\usepackage{bbold}
\usepackage{lipsum}
\usepackage{nccmath}
\usepackage{xcolor}
\hypersetup{colorlinks=true}

\begin{document}

\title{Origins of spontaneous magnetic fields in \texorpdfstring{Sr$_2$RuO$_4$}{Sr2RuO4}}

\author{Yongwei Li}
\email{y.w.li@sjtu.edu.cn}
\affiliation{State Key Laboratory of Micro-nano Engineering Science, Tsung-Dao Lee lnstitute \& School of Physics and Astronomy, Shanghai Jiao Tong University, Shanghai 201210, China}

\author{Rustem Khasanov}
\affiliation{PSI Center for Neutron and Muon Sciences CNM, 5232 Villigen PSI, Switzerland}

\author{Stephen P. Cottrell}
\affiliation{STFC-ISIS Facility, Rutherford Appleton Laboratory, Harwell Campus, Chilton, Oxfordshire, OX11 0QX, United Kingdom}

\author{Naoki Kikugawa}
\affiliation{National Institute for Materials Science, Tsukuba 305-0003, Japan}

\author{Yoshiteru Maeno}
\affiliation{Toyota Riken - Kyoto-Univ. Research Center (TRiKUC), Kyoto 606-8501, Japan}

\author{Binru Zhao}
\affiliation{School of Physics and Astronomy, Shanghai Jiao Tong University, Shanghai 201210, China}

\author{Jie Ma}
\affiliation{School of Physics and Astronomy, Shanghai Jiao Tong University, Shanghai 201210, China}

\author{Vadim Grinenko}
\email{vadim.grinenko@sjtu.edu.cn}
\affiliation{State Key Laboratory of Micro-nano Engineering Science, Tsung-Dao Lee lnstitute \& School of Physics and Astronomy, Shanghai Jiao Tong University, Shanghai 201210, China}

\begin{abstract}
The nature of the broken time reversal symmetry (BTRS) state in Sr$_2$RuO$_4$ remains elusive, and its relation to superconductivity remains controversial. There are various universal predictions for the BTRS state when it is associated with a multicomponent superconducting order parameter. In particular, in the BTRS superconducting state, spontaneous fields appear around crystalline defects, impurities, superconducting domain walls and sample surfaces. However, this phenomenon has not yet been experimentally demonstrated for any BTRS superconductor. Here, we aimed to verify these predictions for Sr$_2$RuO$_4$ by performing muon spin relaxation ($\mu$SR) measurements on Sr$_{2-y}$La$_{y}$RuO$_4$ single crystals at ambient pressure and stoichiometric Sr$_2$RuO$_4$ under hydrostatic pressure. 
We observed that the enhanced muon spin depolarisation rate in the superconducting state $\Delta\lambda$ (in the limit of zero temperature), monotonically decreases with La doping and hydrostatic pressure. The observed behaviour is consistent with $\Delta\lambda\propto T_{\rm c}^2$, indicating that homogeneous La substitution is not a source of spontaneous magnetic fields, but the spontaneous fields are altered by the suppression of superconductivity with La-doping and hydrostatic pressure. Qualitatively different behaviour is observed for the effects of disorder and Ru inclusions. By analysing a large number of samples measured in the previous works, we found that the $\Delta\lambda$ is small for pure Sr$_2$RuO$_4$ single crystals and increases with disorder or impurities. The strongest enhancement is observed in the crystals with Ru-inclusions, which show enhanced $T_{\rm c}$. The comparative study allowed us to conclude that spontaneous fields in the BTRS superconducting state of Sr$_2$RuO$_4$ appear around non-magnetic inhomogeneities and, at the same time, decrease with the suppression of $T_{\rm c}$. The observed behaviour is consistent with the prediction for multicomponent BTRS superconductivity in Sr$_2$RuO$_4$. The results of the work are relevant to understanding BTRS superconductivity in general, as they demonstrate, for the first time, the relationship among the superconducting order parameter, the BTRS transition, and crystal-structure inhomogeneities.
\end{abstract}

\maketitle

As an unconventional superconductor, strontium ruthenate (Sr$_2$RuO$_4$) has been a research hotspot in condensed matter physics for nearly three decades \cite{maeno1994}. Despite extensive research, the structure of its superconducting order parameter and the underlying pairing mechanism remain under intense debate \cite{kivelson2020, sigrist2019}. A key feature of its superconducting state is broken of time-reversal symmetry (BTRS), but its  
relationship to the superconducting order parameter remains controversial.  
The most compelling evidence for the BTRS superconductivity is given by the muon spin rotation/relaxation ($\mu$SR) experiments~\cite{luke1998}, the Polar Kerr effect~\cite{xia2006}, the spontaneous superconducting diode effect~\cite{anwar2023} and some experiments on the Josephson effect measurements~\cite{Fermin2025, Kidwingira2006}. If the BTRS state is intrinsic to superconductivity, the order parameter must be an imaginary two-component. This results in various proposals for the symmetry of the order parameter in Sr$_2$RuO$_4$~\cite{maeno2024, maeno_thirty_2024, leggett2021}. Given a recent revision of the Knight shift measurements~\cite{pustogow2019, ishida2020, chronister2021}, the most favourable possibilities are accidentally degenerated order parameters such as $s+id$~\cite{romer2019}, $d+ig$~\cite{kivelson2020, clepkens2021, Yuan2021}, and out-of-plane chiral state $d_{\rm xz}+id_{\rm yz}$~\cite{suh2020}. The multicomponent nature of the superconducting order parameter is consistent with ultrasound experiments~\cite{benhabib2021, ghosh2021}. However, other measurements cast doubt on whether BTRS is intrinsic to superconductivity. The main concern regarding chiral superconductivity is given by the measurements under uniaxial strain, including the absence of a cusp in the uniaxial strain dependence of the superconducting critical temperature ($T_{\rm c}$)~\cite{hicks2014, watson2018, jerzembeck2022,arXiv:2509.10215} and the missing anomaly at the BTRS transition temperature ($T_{\rm BTRS}$) in the specific heat and the elastocaloric effect~\cite{li2021_pnas, li2022_nature}. 

Regardless, the specific symmetry of the superconducting order parameter, the theoretical models for BTRS superconductivity provide a universal prediction that spontaneous currents can emerge at domain walls, sample surfaces and around inhomogeneities that perturb the phases and amplitudes of the order parameter components \cite{Matsumoto1999,Furusaki2001,Kirtley2007,Lee2009, Garaud2014, Garaud2016, kivelson2020, sigrist2019, etter2018, speight2021}. Whereas the strength and the structure of the fields depend on the details of the superconducting order parameter, electronic band structure and defect properties. 
The zero field (ZF) $\mu$SR, since it is a local probe, is an ideal tool to detect these fields regardless of their structure. 
However, the relationship of the spontaneous fields with disorder has never been experimentally tested in any BTRS superconductor. Mainly because ZF-$\mu$SR measurements are very time-consuming, it is practically unrealistic to obtain sufficient beamtime for a systematic study. Here, we combined our ZF-$\mu$SR measurements on Sr$_{2-y}$ La$_{y}$RuO$_4$ single crystals at ambient pressure and the measurements of Sr$_2$RuO$_4$ under hydrostatic pressure, with a systematic analysis of accumulated literature data over the last 30 years. The comparison of the observed behaviour of the spontaneous magnetic fields with doping, disorder, Ru-inclusions and hydrostatic pressure allowed us to conclude that the behaviour of the ZF-muon spin relaxation rate in Sr$_2$RuO$_4$ is qualitatively consistent with the predictions for the multicomponent BTRS superconductivity. Our data also provide a solution to the puzzle of a missing cusp in the strain dependence of $T_{\rm c}$.

The $\mu$SR experiments were performed on the EMU spectrometer at the ISIS pulsed neutron and muon source. High-quality single crystals of Sr$_{2-y}$La$_{y}$RuO$_4$ ($y=0.01$ and $y=0.04$) were mounted on a copper holder in a dilution refrigerator.

The measurements were performed on the single crystals with the lentgh of about 1 cm and 3mm in diameter cut from the rods grown by the float-zone technique. The crystal was cleaved along the single crystalline $ab$-plane on several rectangular pieces to obtain a large enough sample surface. The muon spin polarization were along the crystallographic $c$-axis. 

The measurements under pressure up to about 1.37GPa were performed at the GPD spectrometer of the SMuS, PSI, in the He3 cryostat using the Cu-beryllium pressure cell specially designed for the measurements with a smaller sample volume. In this case, the muon spin polarization was in the $ab$-plane. For the measurements, we used the Sr$_2$RuO$_4$ sample (rod C171) characterized in the previous experiments under pressure up to 0.95GPa~\cite{grinenko2021natcommun}.

\begin{figure*}[t]
    \centering
    \includegraphics[width=0.95\textwidth]{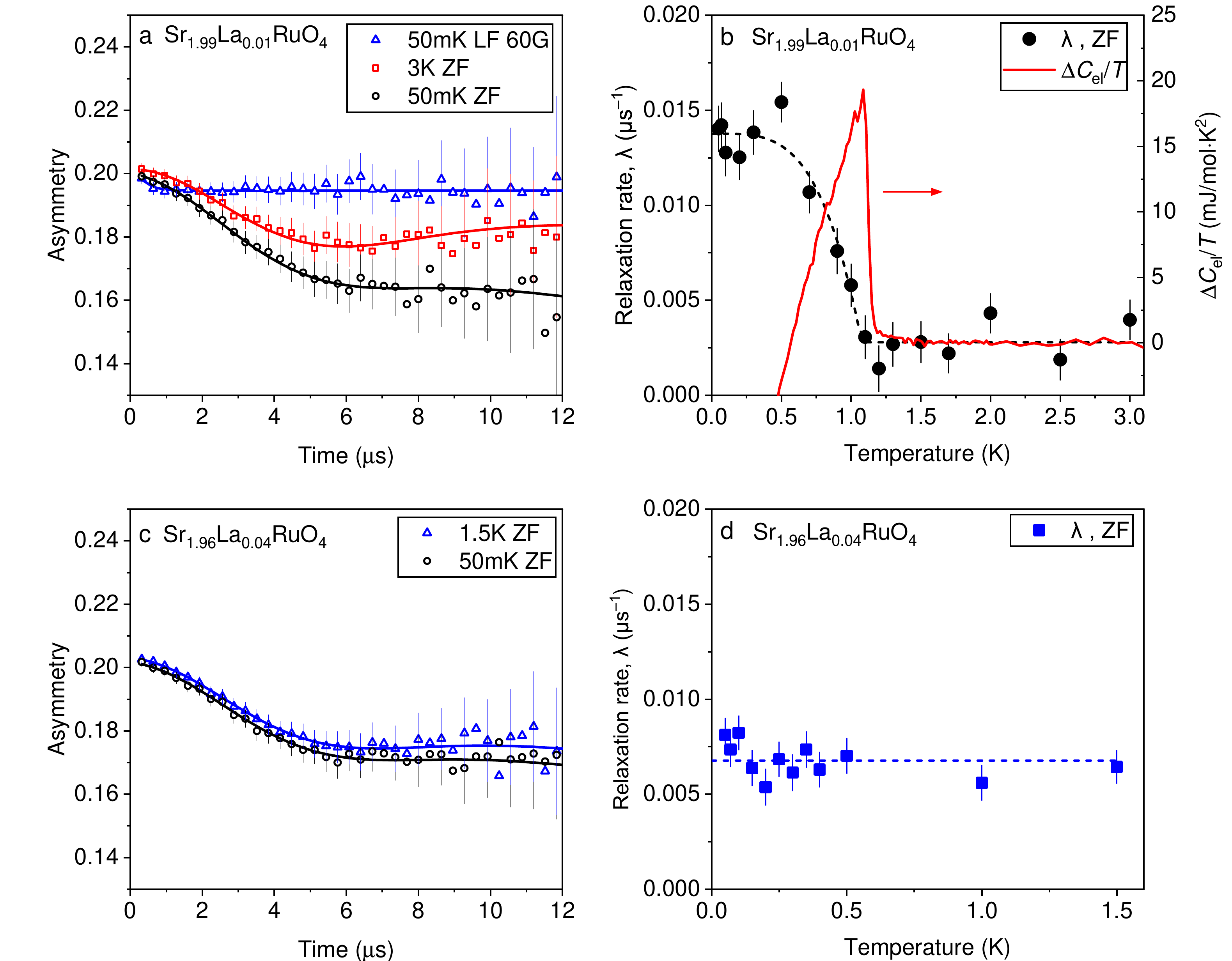}
    \caption{Zero-field $\mu$SR data for Sr$_{2-y}$La$_{y}$RuO$_4$ single crystals. (a) Time evolution of the muon spin asymmetry for $y=0.01$ in the normal (3 K) and superconducting (0.05 K) states. (b) Extracted ZF muon spin relaxation rate $\lambda$ vs. temperature for $y=0.01$. The solid curve represents the best fit to Eq.~(\ref{eq:lambda_T_fit}) with the following parameters: $\lambda_0 = 0.0028(4)~\mu\mathrm{s}^{-1}$, $\Delta\lambda = 0.0110(6)~\mu\mathrm{s}^{-1}$, $n = 3.7(13)$, and $T_{\rm BTRS} = 1.08(5)$~K. The bulk superconducting transition temperature determined from the specific heat measurements is $T_c = 1.13(2)$~K and is consistent with the TF data shown in Fig.~\ref{fig:appendix_A1}. The relaxation rate within the experimental errors is enhanced at $T_{\rm c}$. (c) Time evolution of the muon spin asymmetry for $y=0.04$ at 1.5 K and 0.05 K. (d) Extracted ZF relaxation rate $\lambda$ vs. temperature for $y=0.04$, which is essentially temperature independent in the measured temperature range. In the sample with $y=0.04$, bulk superconductivity is suppressed below 0.05 K as demonstrated in TF measurements (Fig.~\ref{fig:appendix_A1}).}
    \label{fig:fig1}
\end{figure*}

\begin{figure*}[t]
    \centering
    \includegraphics[width=0.95\textwidth]{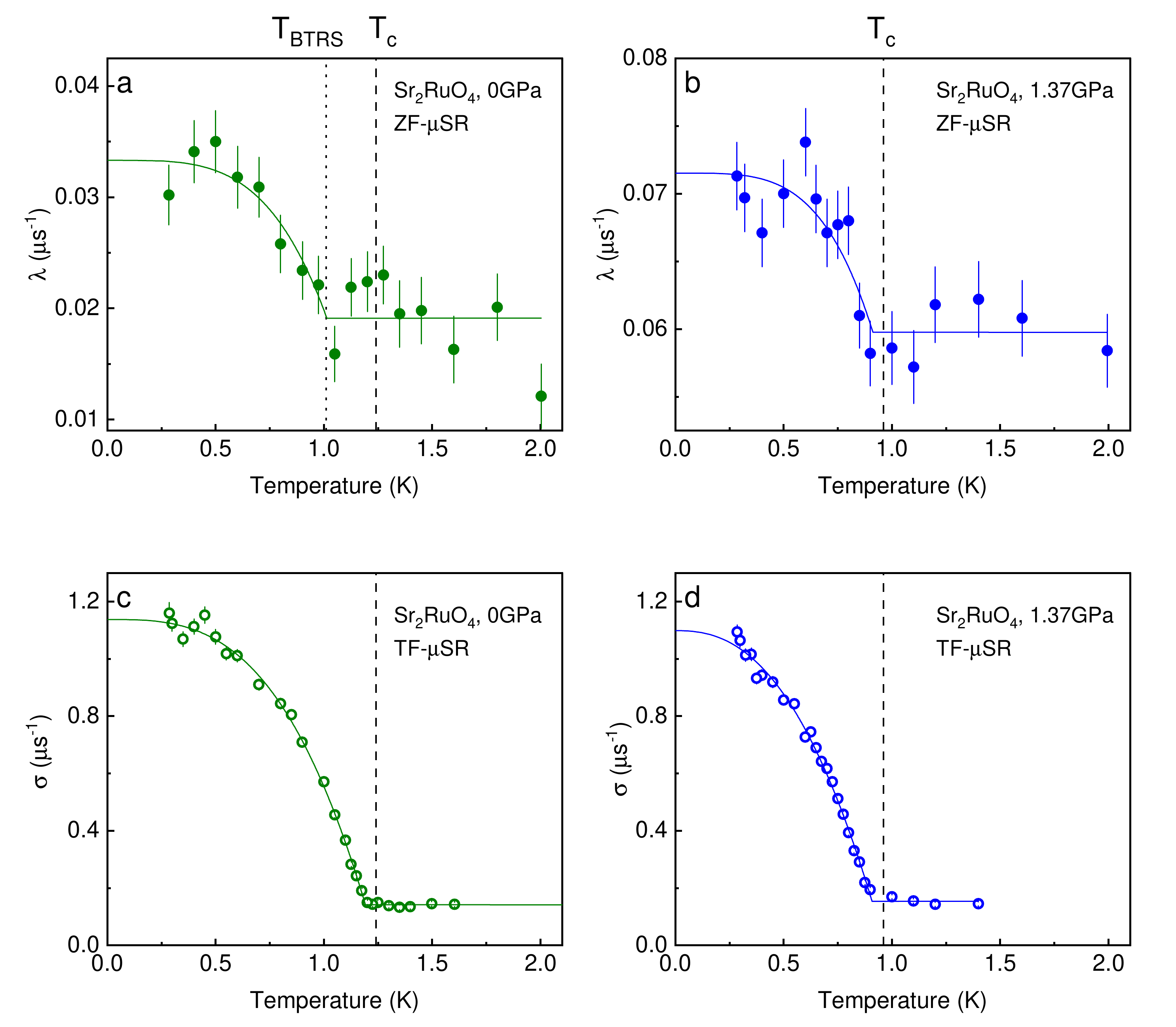}
    \caption{Results of $\mu$SR measurements on Sr$_2$RuO$_4$ under hydrostatic pressure. 
    (a) Zero-field (ZF) muon spin relaxation rate $\lambda$ versus temperature under Zero Pressure (0 GPa). The solid line represents the fit to the phenomenological model Eq.~(\ref{eq:lambda_T_fit}) with parameters: $\lambda_0 = 0.0191(11)~\mu\mathrm{s}^{-1}$, $\Delta\lambda = 0.0142(26)~\mu\mathrm{s}^{-1}$, $n = 3.9(27)$, and $T_{\rm BTRS} = 1.01(7)$~K. The vertical dashed line indicates the superconducting transition $T_c = 1.24(1)$~K determined from TF measurements, showing a splitting between $T_{\rm BTRS}$ (dotted line) and $T_c$. (b) ZF relaxation rate $\lambda$ under High Pressure ($P = 1.37$ GPa). The solid line is the best fit with Eq.~\ref{eq:lambda_T_fit}: $\lambda_0 = 0.0598(11)~\mu\mathrm{s}^{-1}$, $\Delta\lambda = 0.0118(15)~\mu\mathrm{s}^{-1}$, $n = 4$, and $T_{\rm BTRS} = 0.91(4)$~K. The vertical dashed line marks $T_c = 0.97(2)$~K. 
    (c) and (d) Temperature dependence of the transverse-field (TF) Gaussian relaxation rate $\sigma$ at 0 GPa and 1.37 GPa, respectively. These measurements track the formation of the flux-line lattice and were used to determine the intrinsic bulk $T_c$ values indicated by the dashed lines in panels (a) and (b).}
    \label{fig:fig2}
\end{figure*}

We conducted zero-field (ZF), longitudinal-field (LF), and transverse-field (TF) $\mu$SR measurements.
The zero-field (ZF) time spectra of La-doped samples were analysed using a combination of an exponential decay and a static Gaussian Kubo-Toyabe function:

\begin{equation}
    A(t) = A_0 \exp(-\lambda t) + A_{KT} G_{KT}(\Delta, t)
    \label{eq:zf_bg}
\end{equation}

where the first term represents a weak dynamic relaxation background, and the second term, the Kubo-Toyabe function $G_{KT}(\Delta, t) = \frac{1}{3} + \frac{2}{3}(1 - (\Delta t)^2) \exp(-\frac{1}{2}(\Delta t)^2)$, describes the contribution from the nuclear moments of the copper sample holder. $A_{KT}$ was about 10\% of the total asymmetry for both La-doped samples. For measurements under hydrostatic pressure, muon stops in the pressure cell walls contribute about 50\% of the total asymmetry, and the data were analysed using a modified expression for the pressure cell contribution as described in Ref.~\cite{grinenko2021natcommun}. The background contribution in both cases is temperature independent in the measured temperature range and can therefore be reliably separated from the sample signal. The zero field data for La-doped samples at ambient pressure are summarized in Fig.~\ref{fig:fig1}, and the data for non-doped Sr$_{2}$RuO$_4$ under hydrostatic pressure in Fig.~\ref{fig:fig2}.

\begin{figure}[t]
    \centering
    \includegraphics[width=\columnwidth]{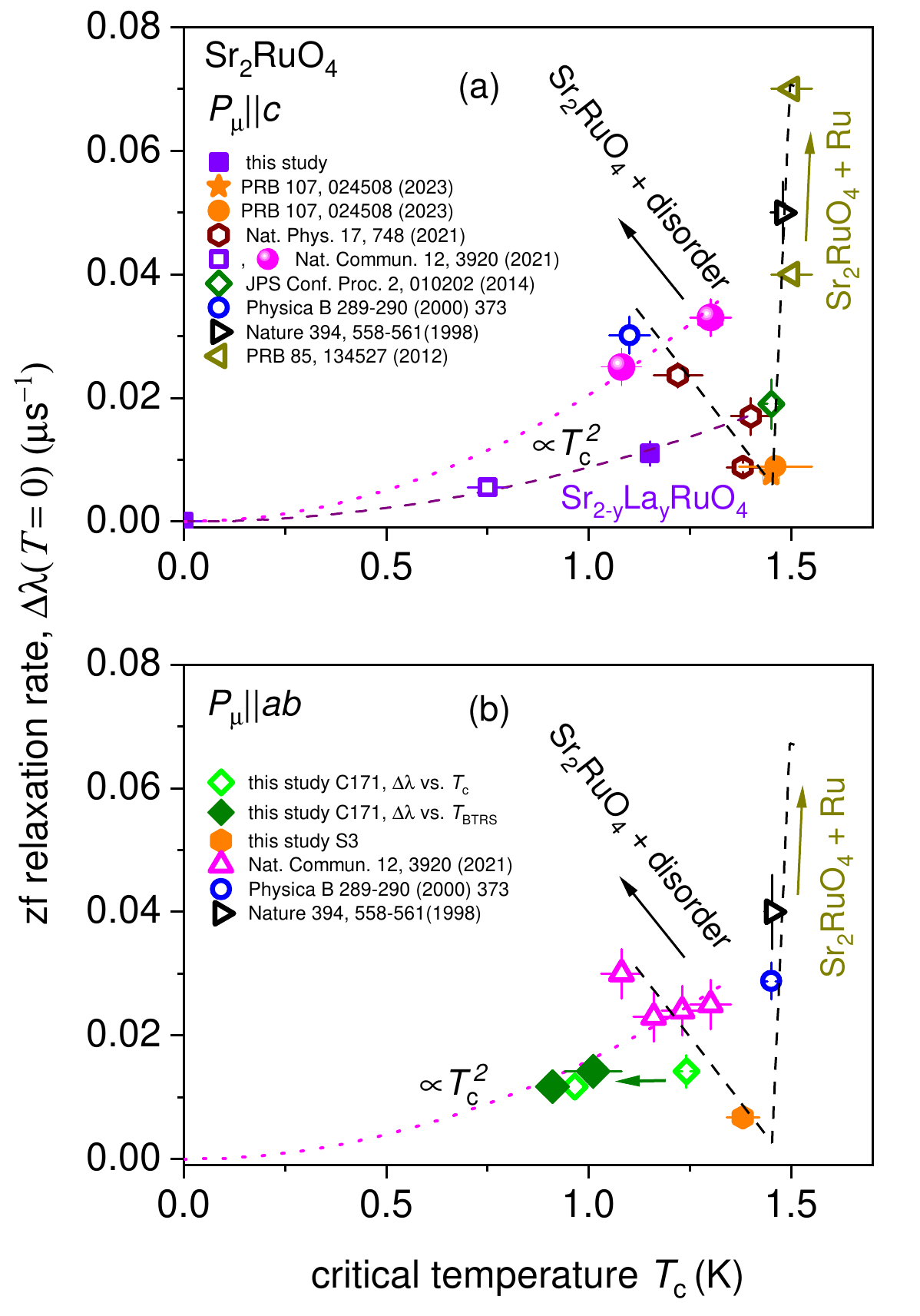}
    \caption{Summary plot of the enhanced muon spin depolarization rate $\Delta\lambda(T=0)$ vs. $T_c$ in Sr$_2$RuO$_4$ and its derivatives. The plot summarizes the data from this study and from the literature. The overall behaviour supports the conclusion that $\Delta\lambda(T=0)\propto nJ_s$ with $n$ inhomogeneity or defect density and $Js$ the strength of spontaneous currents induced by non-magnetic inhomogeneities in the superconducting state with broken time reversal symmetry. For La-doped samples and for Sr$_2$RuO$_4$ under hydrostatic pressure, the density $n$ is nearly unchanged, resulting in a trend $\Delta\lambda \propto nJ_s \propto T_c^2$. In contrast, the samples with Ru inclusions and random disorder (presumably Ru vacancies) show enhanced $\Delta\lambda(T=0)$ dominated by the enhancement of $n$. For further explanations, see the text. The data from the literature  \cite{PhysRevB.107.024508,Nat.Phys.17,grinenko2021natcommun, JPS2014, luke2000unconventional, luke1998, PhysRevB.85.134527}. }
    \label{fig:fig3}
\end{figure}

To extract the BTRS transition temperature and the magnitude of the spontaneous relaxation rate in the superconducting state, the temperature dependence of the relaxation rate $\lambda(T)$ was fitted with the following phenomenological functional form:
\begin{equation}
\lambda (T)=\left\{\begin{array}{ll}{\lambda }_{0}, & T\;> \;{T}_{{\rm{BTRS}}} \\ {\lambda }_{0}+{{\Delta }}\lambda \left[1-{\left(\frac{T}{{T}_{{\rm{BTRS}}}}\right)}^{n}\right], & T\;<\;{T}_{{\rm{BTRS}}} \end{array}\right.
\label{eq:lambda_T_fit}
\end{equation}
where $\lambda_0$ is the normal state relaxation rate above $T_{\rm BTRS}$, and $\Delta\lambda$ is the zero-temperature enhancement of the relaxation rate due to spontaneous magnetic fields. The obtained $\Delta\lambda$ value was then used to plot Fig.~\ref{fig:fig3}.

The transverse-field (TF) spectra are fitted using a function combining an oscillating signal with a non-oscillating background:
\begin{equation}
    A(t) = A_0 \cdot e^{-\frac{1}{2}(\sigma t)^2} \cdot \cos(\omega t + \phi) + A_{bg}
    \label{eq:tf_fit}
\end{equation}
Here, the Gaussian decay rate $\sigma$ 
of the first term includes the sample and T-independent Cu-holder or pressure cell contribution with $\sigma$ given by  $\sigma^2= \sigma^2_{\rm sc} + \sigma^2_{\rm nm}$,
where $\sigma_{\rm sc}$ and $\sigma_{\rm nm}$ are the flux-line lattice and nuclear moment
contributions, respectively. $\sigma_{\rm sc} \propto \lambda_{\rm sc}^{-2}$, where $\lambda_{\rm sc}$ is the superconducting magnetic penetration depth. The temperature dependence of $\sigma$ is used to define $T_{\rm c}$ under hydrostatic pressure. 
The second term, $A_{bg}$ is a time and temperature-independent constant background amounting to about 5\% of the total asymmetry, representing the fraction of muons whose initial spin direction is parallel or anti-parallel to the external field and thus does not contribute to the precession signal.

Figure \ref{fig:fig1} 
summarizes the results of the ZF-$\mu$SR measurements of Sr$_{2-y}$La$_{y}$RuO$_4$ single crystals with $y=0.01$ and $y=0.04$. For $\mathrm{Sr}_{1.99}\mathrm{La}_{0.01}\mathrm{RuO}_4$, the ZF time spectra (Fig. \ref{fig:fig1}a) clearly show an enhanced muon spin relaxation rate in the superconducting state (0.05 K) compared to the normal state (3 K). The weak longitudinal field decouples the muon spins completely, suggesting a static nature of the internal fields.  
The ZF relaxation rate $\lambda$, extracted using Eq. (\ref{eq:zf_bg}), increases sharply below $T_{\rm c} \approx 1.13(2)$~K (Fig. \ref{fig:fig1}b), obtained from the specific heat measurements of the sample part used in the $\mu$SR experiments. The estimated value of $T_{\rm BTRS} \approx 1.08(5)K$ within the errors coincides with $T_{\rm c}$. This behaviour is usually observed in Sr$_2$RuO$_4$, providing clear evidence for the BTRS state in the $y=0.01$ doped sample.

In contrast, the ZF relaxation rate is temperature independent within the experimental error bars for the sample with $y=0.04$ (Figs.~\ref{fig:fig1}c and~\ref{fig:fig1}d). The TF Gaussian decay rate $\sigma$ is also temperature-independent, indicating complete suppression of bulk superconductivity by 2\% of La doping, which is in line with the literature data~\cite{Kikugawa2004} (see supplementary information Fig.~\ref{fig:appendix_A1}). First of all, these observations suggest that the BTRS signal is directly related to the superconducting phase. Another important result is that $T_{\rm BTRS}$ do not split from $T_{\rm c}$ within the experimental errors in Sr$_{2-y}$La$_{y}$RuO$_4$ conforming the previous results for $y=0.02$~\cite{grinenko2021natcommun}.

To further verify the relationship between $T_{\rm c}$ and the spontaneous magnetic fields with a fixed disorder, we performed $\mu$SR measurements on $\mathrm{Sr}_2\mathrm{RuO}_4$ under hydrostatic pressure. Figure \ref{fig:fig2} shows the temperature dependence of the TF Gaussian decay rate $\sigma$ under Zero Pressure (ZP) and $P = 1.37(3) GPa$. Both ZP (Fig. \ref{fig:fig2}c) and $P = 1.37(3) GPa$ (Fig. \ref{fig:fig2}d) data exhibit a clear superconducting transition, marked by the increase in $\sigma$. To accurately determine zero field $T_{\rm c}$, we extrapolated the one obtained from the TF-$\mu$SR measurements to zero field using the experimentally obtained temperature dependence of the upper critical field $H_{\rm c2}$ as shown in the supplementary information Fig.~\ref{fig:appendix_A3}. The $H_{\rm c2}$ for various samples with different $T_{\rm c}$, as well as for Sr$_2$RuO$_4$ under pressure, scales on a single curve, allowing a reliable prediction of zero-field $T_{\rm c}$ using TF measurements. We note that the $T_{\rm c}$ defined in this way for La-doped samples agrees well with specific heat measurements in zero field. 
Surprisingly, we observed that extrapolated to zero field $T_{\rm c}^{\rm ZP}\approx 1.25(2)$K is noticeably higher than the $T_{\rm BTRS}\approx 1.01(7)$K obtained from the fit of ZF-$\mu$SR data. By applying $P = 1.37(3) GPa$ the splitting between transitions is reduced with $T_{\rm c}^{\rm ZP}\approx 1.0(2)$K and $T_{\rm BTRS}\approx 0.91(4)$K. 

The splitting at zero pressure is unexpected since it was not observed in the first experiments reported in Ref.~\cite{grinenko2021natcommun}. The old and new data are compared in Fig.~\ref{fig:appendix_comparison}. It is seen that $T_{\rm c}$ values are nearly the same for the two measurements. The small difference is explained by the fact that for the new measurement, only crystals from the batch C171 were used, while for the old measurements, a larger sample, combined from two different batches (C140 and C171), with similar fractions, was used. The $T_{\rm c}$ difference is consistent with the specific heat data (see the supplementary information in Ref.~\cite{grinenko2021natcommun}). Therefore, a such large change in $T_{\rm BTRS}$ cannot be explained by the observed small difference in $T_{\rm c}$. Also, the data in Fig.~\ref{fig:appendix_comparison} exclude that the crystals from batch C171 had intrinsically split transitions (previously, we observed a large splitting between $T_{\rm c}$ and $T_{\rm BTRS}$ at zero pressure~\cite{grinenko2021_natphys} for one of the Sr$_2$RuO$_4$ samples) since no visible feature can be found around $T_{\rm BTRS}\approx 1.01(7)$K in the first measurements (old data). 

Given that transitions split under uniaxial strain~\cite{grinenko2021_natphys,PhysRevB.107.024508}, we concluded that the splitting of the transitions at zero pressure is likely caused by the residual strain left after releasing pressure after the first experiment. Note, $T_{\rm c}$ is unaffected by this residual strain within the error bars of the measurements. Comparison with the data obtained under uniaxial strain indicates that the residual strain is likely [110] uniaxial or pure $B_{2g}$ shear strain, where a negligibly small effect on $T_{\rm c}$ was observed~\cite{science.1248292, arXiv:2509.10215}. In contrast, the effect of [100] strain on $T_{\rm c}$ is much more pronounced~\cite{science.1248292, grinenko2021_natphys}. The strong sensitivity of $T_{\rm BTRS}$ to [110] strain was recently suggested from the ZF-$\mu$SR study \cite{PhysRevB.107.024508}. It is worth emphasising that the observed strong sensitivity of $T_{\rm BTRS}$ to residual strain without noticeable effect on $T_{\rm c}$ is a quite important observation and it can explain the splitting of the transitions observed for some other samples at zero applied strain~\cite{grinenko2021_natphys}. Moreover, it may solve the problem of the missing cusp in the strain dependence of $T_{\rm c}$ expected for BTRS superconductors with $T_{\rm c}=T_{\rm BTRS}$ ~\cite{hicks2014, watson2018, jerzembeck2022,arXiv:2509.10215} if we assume that transitions split at least slightly for all studied samples. Indeed, real samples inevitably have some small residual strain due to stacking faults, dislocations, vacancies and various inclusions such as Ru and Ru-based phases. Theoretically, it was shown that even a small splitting is enough to eliminate the cusp~\cite{PhysRevB.87.144511, arXiv:2509.19137} and to result in a parabolic dependence of $T_{\rm c}$ on the uniaxial strain observed in the experiment for [100] direction~\cite{science.1248292, science.aaf9398, PhysRevB.98.094521}.                   

Figure \ref{fig:fig3} summarizes the results by plotting the zero-temperature relaxation rate increment $\Delta\lambda(T=0) = \lambda(T=0) - \lambda(T> T_{\rm c})$ against $T_c$ for a wide range of samples, $\lambda(T=0)$ is the extrapolated relaxation rate to the limit of zero temperature. The central conclusion of this paper is that $\lambda(T=0)$ behaviour of various samples is consistent within a magnetic response expected for multicomponent superconductivity. The spontaneous magnetic fields in BTRS superconductors are generated if the phase or amplitude of the multicomponent superconducting order parameter $\Psi=(\Psi_{\rm 1}, \Psi_{\rm 2})$ is perturbed, with $\Psi_{\rm i} = |\Delta_{\rm i}|\exp{(-i\psi_i)}$. In this case, the spontaneous magnetic fields are related to spontaneous supercurrents ($J_s$) induced around inhomogeneities or superconducting domain walls. It can be shown that the induced supercurrents are proportional to a combination of $|\Psi_{\rm i}|^2$ and $|\Psi_{\rm 1}\Psi_{\rm 2}|$. For example, for $s+id$ case, to the linear order in variation of the phase ($\phi(r)$) and amplitude ($\delta_i$) the current $J_s$ is given by $J_s = \sum_i\frac{2e\hbar}{m_i}|\Psi_{i0}^2|(\nabla\phi_i+2\delta_i\nabla\phi_i)-\frac{4e^2}{c}\sum_i\frac{|\Psi_{i0}^2|}{m_i}{\bf A}
+\frac{\hbar e}{m_c}|\Psi_{10}\Psi_{20}|(\hat{x}\partial_x+\hat{y}\partial_y)(\delta_2-\delta_1)$, where the variation of the order parameter around defects is approximated by $\Psi_{\rm i} =\Psi_{\rm i0}[1+\delta_i(r)+i\phi(r)]$, $A$ is a vector potential and $m_c$ parametrizes the interband coupling strength~\cite{Lin2016}. 

It is generally expected that $\Delta\lambda(T=0)$ measured in ZF-$\mu SR$ experiments is proportional to the strength of the spontaneous currents $J_s$ and the number of field sources $n$, i.e. the defect density in the sample volume $\Delta\lambda(T=0)\propto nJ_s$. This behaviour is indeed consistent with Fig.~\ref{fig:fig3}. The relaxation is minimal for clean samples, with $T_{\rm c}\approx 1.4$ K, and largest in samples with Ru inclusions. The Ru-inclusions are nanoparticles which induce a strain field in  Sr$_2$RuO$_4$ and locally enhance $T_{\rm c}$. Several Sr$_2$RuO$_4$ samples in Fig.~\ref{fig:fig3} exhibit reduced $T_{\rm c}$ and a stronger relaxation rate than clean samples. The reduction of $T_{\rm c}$ is expected due to Ru vacancies. In this case, $\Delta\lambda(T=0)$ is smaller compared to the samples with Ru-inclusions, which can be attributed to a weaker perturbation potential and to the suppression of the superconducting order parameter by Ru vacancies, i.e. $J_s \propto\Delta^2\propto T_{\rm c}^2$. 

Remarkably, both La-doped crystals and stochiometric Sr$_2$RuO$_4$ under pressure show a similar decreasing trend. A weakening of the spontaneous fields is expected for hydrostatic pressure, since the defect density and its strength would not be noticeably affected while $T_{\rm c}$ is suppressed monotonously. However, $\Delta\lambda(T=0)$ values under hydrostatic pressure reported in Ref.~\cite{grinenko2021natcommun} noticeably deviate from the data obtained in the present study on a part of the sample from Ref.~\cite{grinenko2021natcommun}, see Fig.~\ref{fig:fig3}b. We found that both new and old data follow the same trend, $\Delta\lambda(T=0) \propto T_{\rm c}^2$, when, for new measurements, we plot the relaxation rate vs. $T_{\rm BTRS}$ instead of $T_{\rm c}$. This observation can be reconciled with the superconducting nature of the BTRS state, assuming that defect density is nearly unchanged (long-range residual strain field) and $\Delta\lambda(T=0)\propto J_s$ is dominated by the specific order parameter component, sensitive to the residual strain, while the second component responsible for $T_{\rm c}$ remains unaffected by the strain. Thus, the observed behaviour of $\Delta\lambda(T=0)\propto T_{\rm BTRS}^2$ is suggestive of an accidentally degenerated order parameter, such as discussed above, an $s+id$ state, since for a chiral state with equal components we would expect both components to contribute to the measured signal. We note that all other samples listed in Fig.~\ref{fig:fig3} didn't show noticeable splitting between $T_{\rm BTRS}$  and $T_{\rm c}$. 

The similar behaviour of $\Delta\lambda(T=0)$ under pressure and with La-doping indicates that, despite the strong pair-breaking effect, La atoms do not induce spontaneous magnetic fields. This can be explained by the homogeneous distribution of La atoms, leading to a relatively uniform suppression of the superconducting order parameter over the sample volume, presumably due to a large superconducting coherence length of Sr$_2$RuO$_4$. We found that uniform suppression of the order parameter is consistent with the specific heat data shown in Fig.~\ref{fig:appendix_A2}. Despite the strong suppression of $T_{\rm c}$ by a tiny amount of La, the superconducting transitions remain sharp for all studied La doping levels. Finally, we concluded that the overall behaviour of the ZF $\Delta\lambda(T=0)$ shown in  Fig.~\ref{fig:fig3} is consistent with the multicomponent nature of the BTRS superconducting order parameter of Sr$_2$RuO$_4$.     

Furthermore, we analysed the anisotropy of the spontaneous fields using data from Sample S3 (see supplementary information Fig. \ref{fig:appendix_A4}). We found that the ratio of the zero-temperature relaxation rates $\Delta\lambda_{ab} / \Delta\lambda_c \approx 0.77(13)$ for this clean sample. This indicates a stronger spontaneous field component aligned in the crystallographic $ab$-plane. Moreover, comparing the data in panels a and b of Fig.~\ref{fig:fig3} indicates that the anisotropy of 0.8 is approximately constant across various samples. This universality in the anisotropy ratio suggests a robust and consistent structure for the magnetic field distribution in the BTRS state of $\mathrm{Sr}_2\mathrm{RuO}_4$ regardless of the type of inhomogeneities. Note, this anisotropy is different compared to the anisotropy of the spontaneous fields found for the Ba$_{\rm 1-x}$K$_{\rm x}$Fe$_2$As$_2$ system~\cite{grinenko_superconductivity_2020}, which indicates a different structure of the superconducting order parameter in this system. According to calculations, the anisotropy of the spontaneous fields depends strongly on the symmetry of the BTRS order parameter and the anisotropy of the Fermi velocities~\cite{grinenko_superconductivity_2020, PhysRevB.98.104504}. In particular, it was found that the dipolar spontaneous fields around defects are quite isotropic for an $s+id$ in contrast to an $s+is$ state. For a chiral superconducting state, the spontaneous fields are expected to be relatively isotropic due to the large spontaneous current loops induced by the chiral order parameter along the domain walls. These current loops would induce a stray field with components in all directions.     

In summary, the analysis of a large number of samples measured in our previous work, as well as reported by other groups, allowed us to conclude that the overall disorder and impurity effect on the relaxation rate in the superconducting state of $\mathrm{Sr}_2\mathrm{RuO}_4$ is consistent with a multicomponent superconducting order parameter that breaks time-reversal symmetry. We also observed that residual strain after hydrostatic pressure measurements split $T_{\rm BTRS}$ and $T_{\rm c}$ transition temperatures. Without noticeable effect on $T_{\rm c}$. This surprising sensitivity of the $T_{\rm BTRS}$ transition to internal strain suggests that the splitting within experimental errors (large in the case of $\mu$SR experiments) may be present in many other samples. This would, for example, explain the surprising lack of a cusp in the uniaxial strain dependence of $T_{\rm c}$. Finally, the behaviour of the $\Delta\lambda$ in the samples with split transitions is suggestive of an accidentally degenerated order parameter in $\mathrm{Sr}_2\mathrm{RuO}_4$.

\section*{Acknowledgment}

This work is supported by the National Natural Science Foundation of China (NSFC) (Grants 12374139, 12350610235 and 12334008), and JSPS KAKENHI (Grant Nos. JP18K04715, JP21H01033, JP22K19093, and 24K01461). We acknowledge the STFC ISIS Neutron and Muon Source for the provision of muon beamtime on the EMU spectrometer. We also thank the Paul Scherrer Institute (PSI) for providing muon beamtime at the Swiss Muon Source (S$\mu$S).

\bibliography{main}

\clearpage

\onecolumngrid


\vspace{2em} 
\begin{center}
    \large \textbf{Supplementary information: Origins of spontaneous magnetic fields in \texorpdfstring{Sr$_2$RuO$_4$}{Sr2RuO4}}
\end{center}
\vspace{0.5em}

\setcounter{equation}{0}
\setcounter{figure}{0}
\setcounter{table}{0}
\makeatletter
\renewcommand{\theequation}{S\arabic{equation}}
\renewcommand{\thefigure}{S\arabic{figure}}
\renewcommand{\thetable}{S\arabic{table}}
\makeatother

\begin{figure*}[b]
    \centering
    \includegraphics[width=0.95\textwidth]{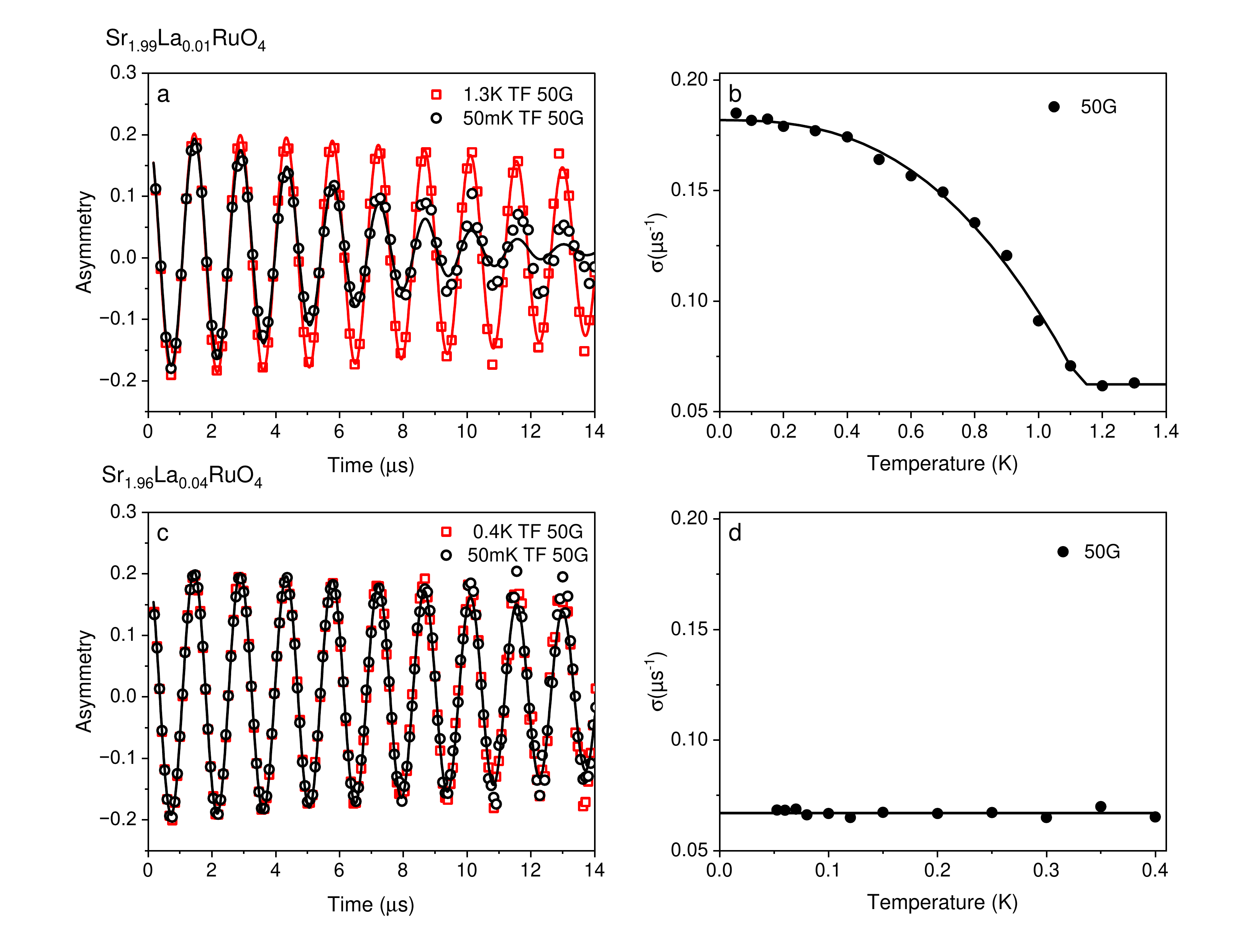}
    \caption{TF $\mu$SR data for Sr$_{2-y}$La$_{y}$RuO$_4$. (a) Time evolution of the muon spin asymmetry for $y=0.01$ under $TF=50G$. (b) Gaussian decay rate $\sigma$ vs. temperature for $y=0.01$, showing an increase below $T_c = 1.13(2)$~K. (c) Time evolution of the muon spin asymmetry for $y=0.04$ under $TF=50G$. (d) $\sigma$ vs. temperature for $y=0.04$. The rate remains constant, demonstrating the suppression of bulk superconductivity.}
    \label{fig:appendix_A1}
\end{figure*}

\section{Transverse Field $\mu$SR Data}
Here we present the Transverse Field (TF) $\mu$SR data, which are primarily used to verify bulk superconductivity and to extract the magnetic penetration depth.

Figure \ref{fig:appendix_A1} shows the temperature dependence of the Gaussian decay rate $\sigma$ for both La-doped samples. For the $y=0.01$ sample (Fig. \ref{fig:appendix_A1}b), $\sigma$ increases sharply below $T_c = 1.13(2)$~K, consistent with the formation of a bulk flux-line lattice. Conversely, for the $y=0.04$ sample (Fig. \ref{fig:appendix_A1}d), $\sigma$ remains constant down to the lowest measured temperature, validating the zero-field findings that both the bulk superconducting signal and the BTRS state are suppressed entirely by the high La-doping.

\section{Specific Heat of Sr$_{2-y}$La$_{y}$RuO$_4$}

\begin{figure}[h]
    \centering
    \includegraphics[width=0.5\columnwidth]{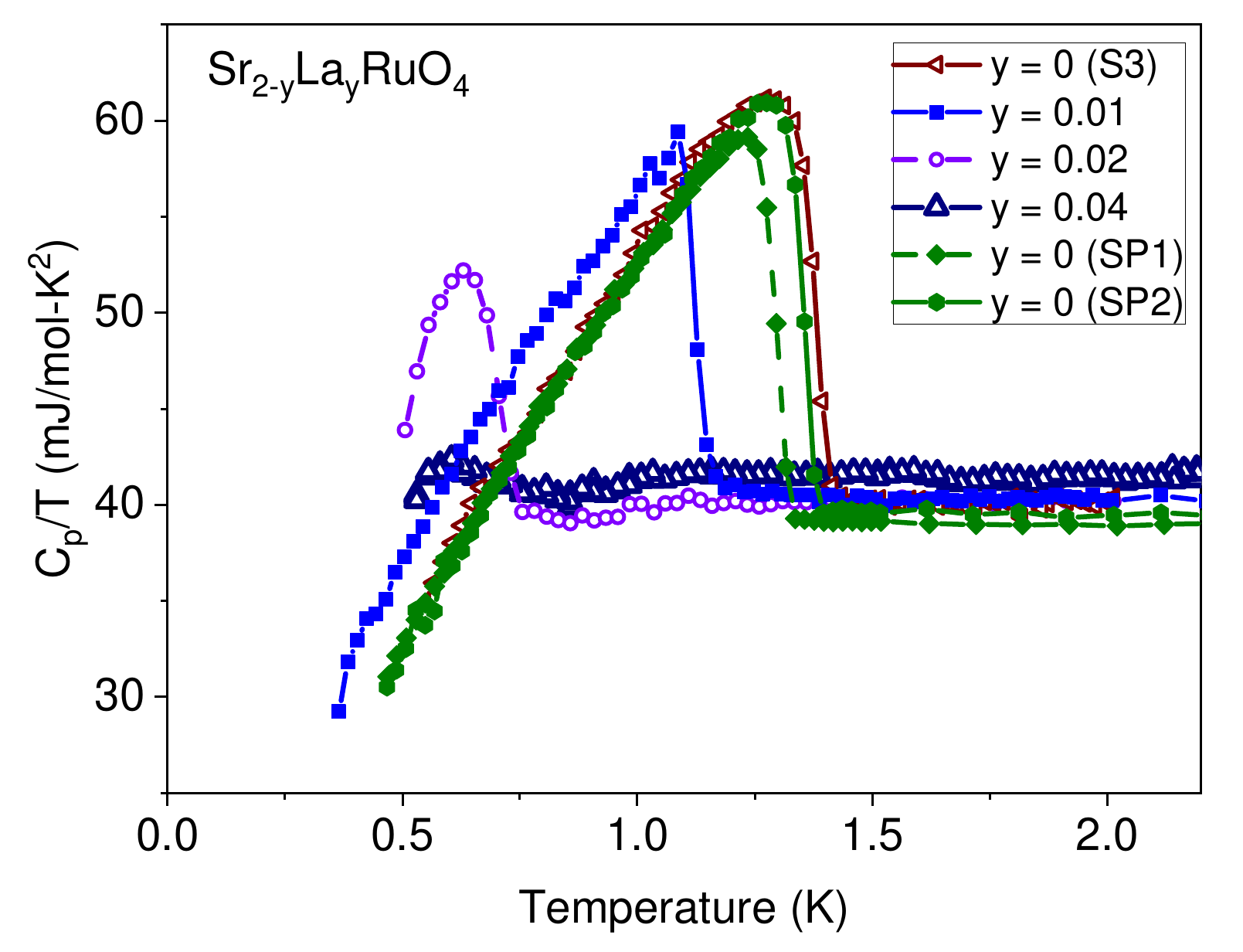}
    \caption{Temperature dependence of the specific heat, plotted as $C_p/T$, for Sr$_{2-y}$La$_{y}$RuO$_4$ single crystals. Superconductivity is suppressed homogeneously with La doping.}
    \label{fig:appendix_A2}
\end{figure}

In the main text, we argue that La substitution suppresses superconductivity homogeneously without inducing local phase and amplitude gradients. To support this claim, we present the specific heat data in Fig.~\ref{fig:appendix_A2}. An anomaly at the superconducting transition is suppressed monotonically toward lower temperatures as La concentration increases, without noticeable broadening, in accord with the assumed homogeneous distribution of La atoms.

\section{$T_c$ Correction under Magnetic Field}
\begin{figure}[h]
    \centering
    \includegraphics[width=0.5\columnwidth]{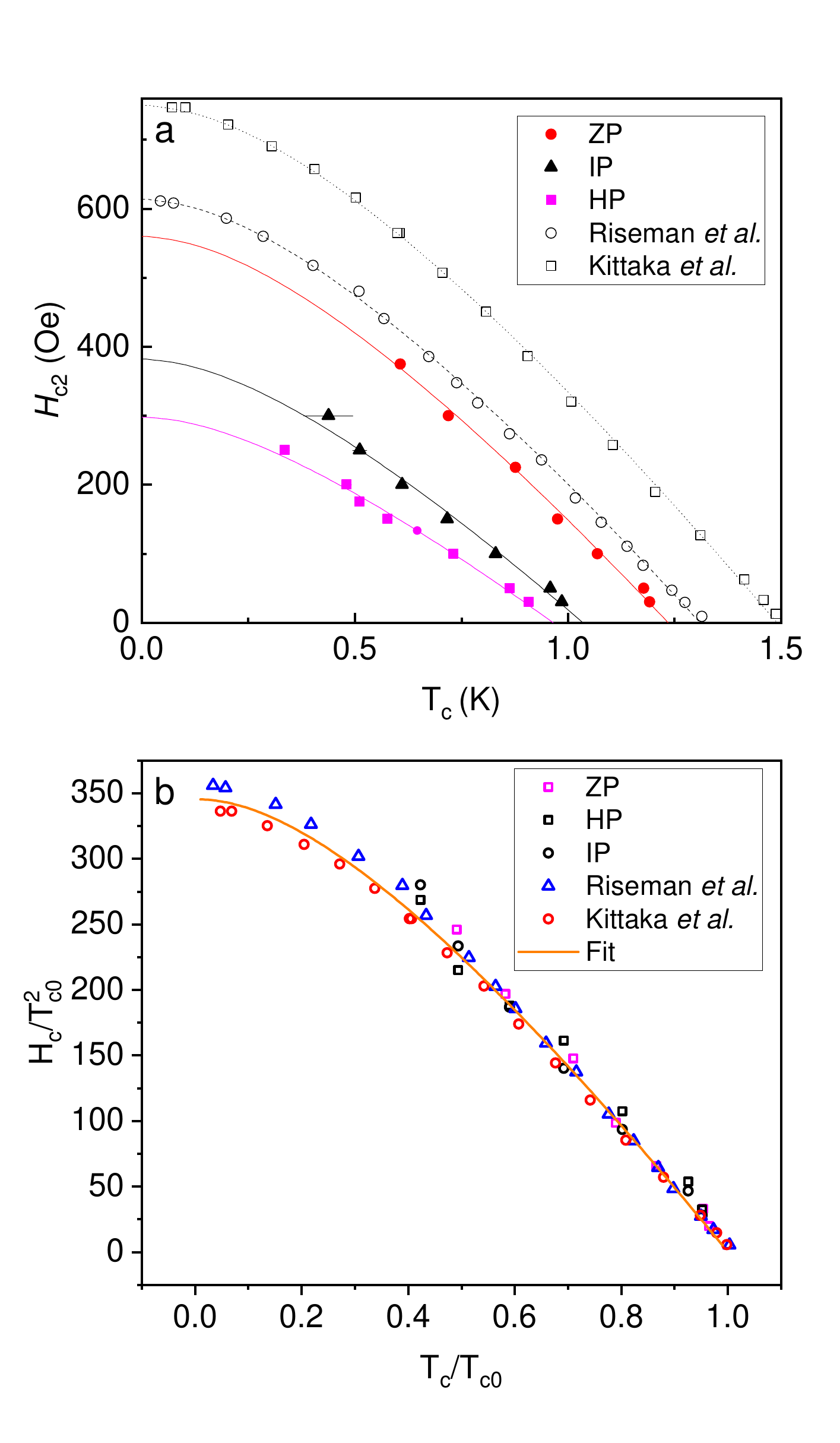}
    \caption{(a) Superconducting upper critical field $H_{c2}$ vs $T$ for Sr$_2$RuO$_4$ samples from the TF - $\mu$SR measurements performed in this study (ZP - zero pressure, IP - intermediate pressure, HP - highest pressure reached in the experiment) and the data from the literature~\cite{riseman_1998,Kittaka2009}. (b) Normalized $H_{c2}/T_{c0}^2$ vs normalized critical temperature $T_c/T_{c0}$. The data collapse onto a universal curve described by the orbital-limited WHH model.}
    \label{fig:appendix_A3}
\end{figure}

To accurately plot the relation between $\Delta\lambda$ and $T_c$ in Fig. \ref{fig:fig3}, we must correct the $T_c$ values obtained from Transverse Field (TF) $\mu$SR measurements to their actual zero-field values ($T_{c0}$). To investigate the behaviour of the upper critical field $H_{c2}$ under hydrostatic pressure, we performed TF measurements at several applied magnetic fields. The obtained experimental data are plotted in Fig. \ref{fig:appendix_A3}a. It is seen that $H_{c2}$ vs. $T_c$ for our samples under pressure and Sr$_2$RuO$_4$ samples from the literature~\cite{riseman_1998,Kittaka2009} follow the Werthamer-Helfand-Hohenberg (WHH) theoretical curves with negligible Pauli paramagnetic effects:
\begin{equation} \label{eq:whh_reduced}
    \ln\left(\frac{T_c}{T_{c0}}\right) = \psi\left(\frac{1}{2} + \frac{C \cdot (H_{c2}/T_{c0}^2)}{T_c/T_{c0}}\right) - \psi\left(\frac{1}{2}\right),
\end{equation}
where $\psi(x)$ is the digamma function, and $C \approx 4.06 \times 10^{-4}$ is the fitting parameter derived from the universal curve. Note that the generalized WHH expansion (involving summation over Matsubara frequencies) simplifies to this analytical form when the Maki parameter $\alpha$ and spin-orbit scattering $\lambda_{so}$ are negligible. This relation allows us to develop a simple procedure to determine a zero-field $T_{\rm c0}$ using a $T_{\rm c}$ value measured in a single applied TF field. To demonstrate this we plotted the normalized upper critical field curves $H_{c2}/T_{c0}^2$ versus the normalized temperature $T_c/T_{c0}$ in Fig. \ref{fig:appendix_A3}b.

\section{Anisotropy in Sample S3}
\begin{figure}[h]
    \centering
    \includegraphics[width=0.5\columnwidth]{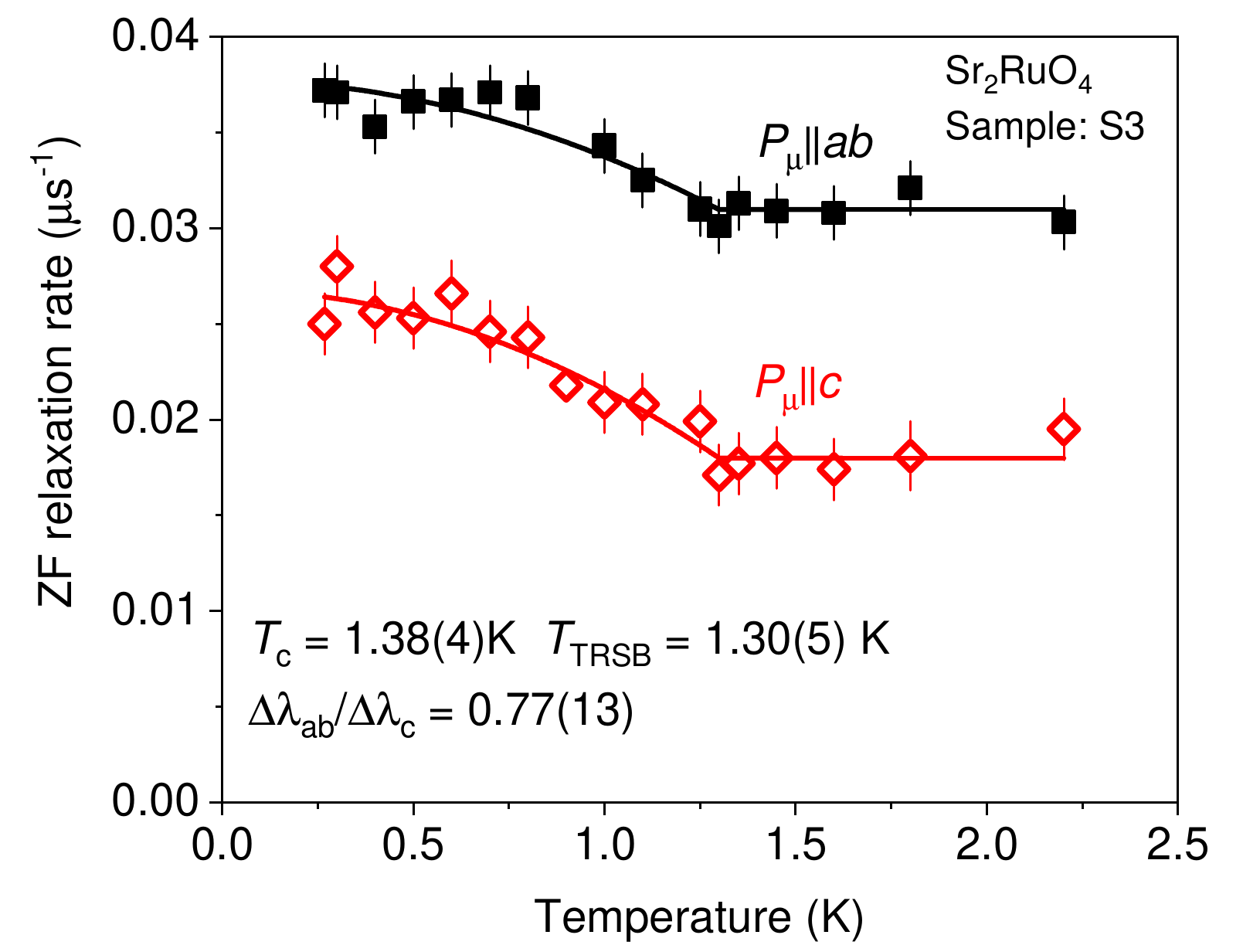}
    \caption{ZF relaxation rate for Sample S3 showing the temperature dependence of the relaxation rate for muon spin polarization parallel to the ab-plane ($P_{\mu} \parallel ab$) and c-axis ($P_{\mu} \parallel c$). The enhancement of the relaxation rate $\Delta\lambda$ below $T_{\rm c}$ is anisotropic.}
    \label{fig:appendix_A4}
\end{figure}

Figure \ref{fig:appendix_A4} shows the zero-field relaxation rate for Sample S3 from Fig.~\ref{fig:fig3}. The measurements reveal an anisotropy in the relaxation rate, with the muon spin polarization parallel to the ab-plane ($P_{\mu} \parallel ab$) showing a weaker signal than parallel to the c-axis ($P_{\mu} \parallel c$). Based on these data, we calculated the ratio $\Delta\lambda_{ab} / \Delta\lambda_c \approx 0.77(13)$. Analyzing the data shown in  Fig.~\ref{fig:fig3} we found that the anisotropy value of about 0.8 holds for most of the samples reported in the literature.

\section{Comparison with Previous $\mu$SR Results under Pressure}

\begin{figure}[h]
    \centering
    \includegraphics[width=0.5\columnwidth]{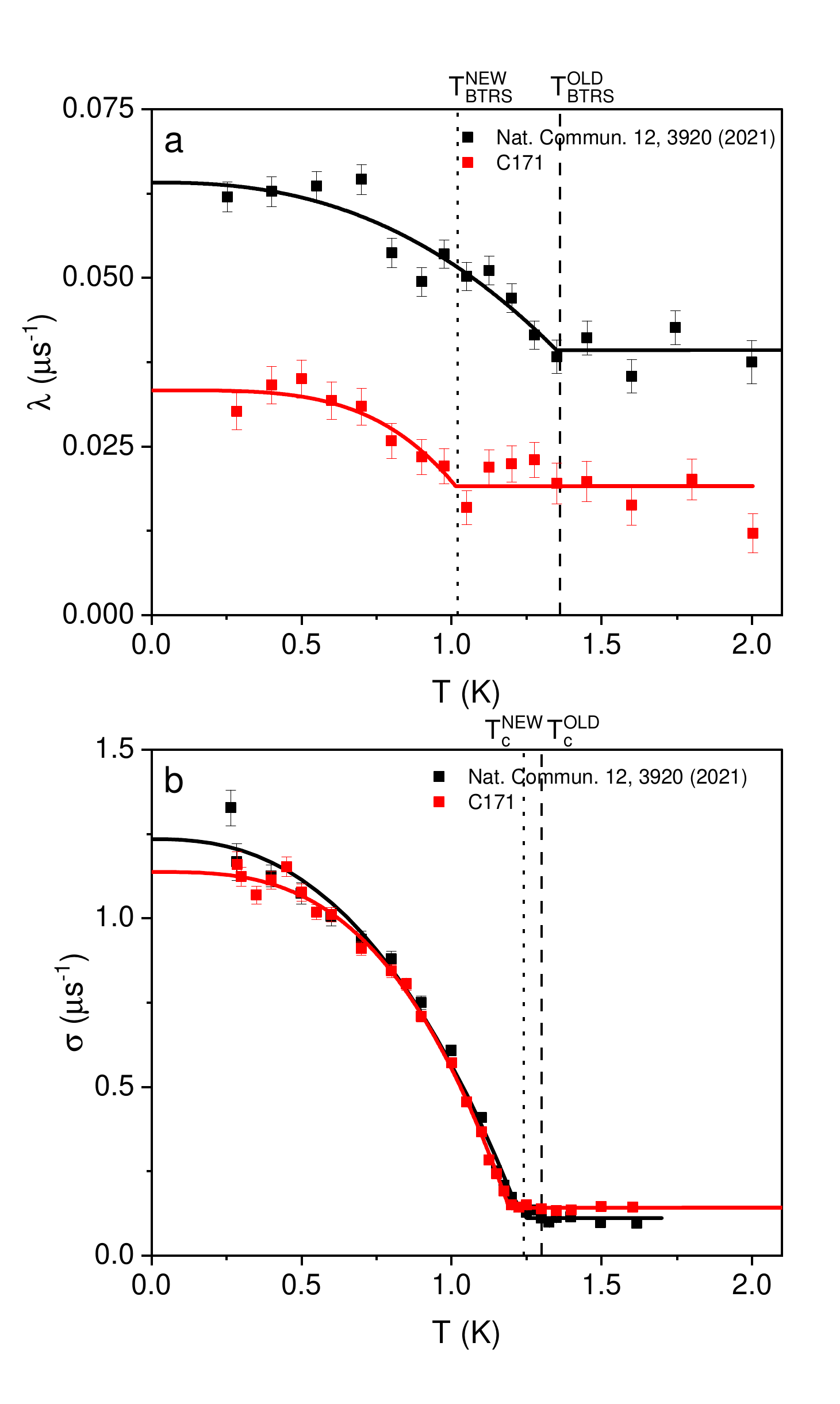} 
    \caption{Comparison of Zero-Field (ZF) and Transverse-Field (TF) $\mu$SR data for the Sr$_2$RuO$_4$ at zero pressure measured inside the pressure cell, contrasting new measurements of the crystals from batch C171 with previous (old) results from Ref.~\cite{grinenko2021natcommun} where crystals from two batches, C140 and C171, were used. For the new measurements, we used the same crystals as in Ref.~\cite{grinenko2021natcommun}, i.e., after the application of hydrostatic pressure. (a) The ZF muon spin relaxation rate $\lambda(T)$ marks the onset of the BTRS state. The transition temperatures are determined as $T^{\text{OLD}}_{\text{BTRS}} = 1.35(25)$~K and $T^{\text{NEW}}_{\text{BTRS}} = 1.01(7)$~K. (b) The TF relaxation rate $\sigma(T)$ reflects the bulk superconducting transition $T_c$. The corresponding critical temperatures extrapolated to zero field, as shown in Fig.~\ref{fig:appendix_A3}, are $T^{\text{OLD}}_{c} = 1.30(5)$~K and $T^{\text{NEW}}_{c} = 1.24(1)$~K indicated by the vertical dashed lines. The featureless behaviour of the old ZF-data around $T^{\text{NEW}}_{\text{BTRS}}$ indicates that, initially, the BTRS and superconducting transition temperatures in the C171 sample were unsplit within experimental error.}
    \label{fig:appendix_comparison}
\end{figure}

To further clarify the nature of the BTRS transition and its relation to the superconducting transition ($T_c$), we compare in Fig.~\ref{fig:appendix_comparison} our current results with the data previously reported in Ref.~\cite{grinenko2021natcommun}. The previous study primarily focused on the combined analysis of the crystals from two batches (C140 and C171). However, in the current measurements, only the crystals from batch C171 were used due to the smaller size of the pressure cell. The comparison revealed that splitting between $T_{\rm BTRS}$ and $T_{\rm c}$ was unlikely to be present in the first measurements. Therefore, we concluded that the splitting was most likely caused by residual strain that remained after pressure was released following the first measurements.



\end{document}